\documentclass[aps,preprint,amsmath,amssymb]{revtex4}

\usepackage{amsmath}
\usepackage{graphicx}
\usepackage{amssymb}

\begin{document}

\newcommand{\ms}{\scriptscriptstyle}

\def\mathbi#1{\textbf{\em #1}}

\title{Biased imitation in coupled evolutionary games in interdependent networks}

\author{M. D. Santos}
\affiliation{Department of Physics \& I3N, University of Aveiro, 3810-193 Aveiro, Portugal}

\author{S. N. Dorogovtsev}
\affiliation{Department of Physics \& I3N, University of Aveiro, 3810-193 Aveiro, Portugal}

\affiliation{A. F. Ioffe Physico-Technical Institute, 194021 St. Petersburg, Russia}
\author{J. F. F. Mendes}

\affiliation{Department of Physics \& I3N, University of Aveiro, 3810-193 Aveiro, Portugal }

\date{\today}

\begin{abstract}
We explore the evolutionary dynamics of two games---the Prisoner's Dilemma and the Snowdrift Game---played within distinct networks (layers) of interdependent networks. 
In these networks imitation and interaction between individuals of opposite layers is established through interlinks. We explore an update rule in which revision of strategies is a {\it biased imitation process}: individuals imitate neighbors from the same layer with probability $p$, and neighbors from the second layer with complementary probability $1-p$. We demonstrate that a small decrease of $p$ from $p=1$ (which corresponds to forbidding strategy transfer between layers) is sufficient to promote cooperation in the Prisoner's Dilemma subpopulation. This, on the other hand, is detrimental for cooperation in the Snowdrift Game subpopulation. We provide results of extensive computer simulations for the case in which layers are modelled as regular random networks, and support this study with analytical results 
for coupled well-mixed populations.
\end{abstract}

\maketitle

Network science has registered tremendous breakthroughs in the last years. Among these findings are the development of analytical and computational tools to study large real networks, as well as the development of more realistic models. These studies interpreted real networks mostly as single, isolated entities. However, the deeper understanding of complex networks is showing that in fact they are generally organized as {\it networks of networks}, and therefore the main focus of scientific research is shifting to {\it multiplex and interdependent networks} \cite{buldyrev2010catastrophic}. 

Interdependent networks are organized in two or more layers (subnetworks), in which the functioning of a node in one layer depends on neighbor nodes in other layers. Multiplex networks \cite{PhysRevLett.111.058701,cozzo2013contact}, on the other hand, are composed of a single type of node and several types (``colors'') of links. One example of a real multiplex network is the transportation network of a city: nodes represent locations in the city, and links are of different type depending on the transportation system(s) that connect two given locations (metro, autobus, etc). Interdependent networks have already been applied in studies in diverse areas of science \cite{helbing2013globally}. There is significant progress in the understanding of their percolation properties and robustness \cite{baxter2012avalanche,parshani2010interdependent,son2012percolation,gao2011robustness,zhou2012assortativity,li2012cascading,buldyrev2010catastrophic,huang2011robustness,vespignani2010complex,radicchi2013abrupt}. Furthermore, interdependent networks 
were also applied in the evaluation of seismic risk \cite{poljanvsek2012seismic,duenas2007seismic}, in the emergence of creativity \cite{csermely2012appearance} and on the impact of such population structure on voting outcome \cite{halu2013connect}.

Network science \cite{dorogovtsev2002evolution,dorogovtsev2003evolution,albert2002statistical}, together with evolutionary game theory \cite{sigmund2010calculus}, helps to understand how cooperative behavior can emerge and evolve in structured populations of selfish individuals. Cooperators contribute with a cost so that another individual can receive a certain benefit. Defectors on the other hand do not contribute, yet reap the benefit (which we assume to be larger than the cost). Interactions between individuals are traditionally modelled by one-shot symmetric two-person games, including the Prisoner's Dilemma (PD) \cite{axelrod1981evolution,rapoport1965prisoner}, the Snowdrift Game (SG) \cite{sugden1986economics} or the Stag-Hunt game (SH) \cite{skyrms2004stag}. In structured populations individuals only interact with their nearest neighbors, and it becomes possible for cooperators to escape exploitation by defectors by forming clusters in which they support each other. This process is known as {\it network reciprocity} \cite{nowak1992evolutionary,nowak2006five,santos2005scale,santos2008social,szabo2007evolutionary,hauert2004spatial,ohtsuki2006simple,roca2009effect}. 

Population structure for games was traditionally modelled as a single, isolated network. Only very recently research has started to evaluate the impact of more complex, interdependent networks on the evolution of cooperation. To the best of our knowledge, all previous works have implemented the same game in all layers \cite{gomez2012evolution,wang2013interdependent,gomez2012evolutionary,wang2012probabilistic,jiang2013spreading,wang2013optimal,wang2012evolution}. Different layers however can represent distinct environments, characterized by their game rules, which can interact or be coupled. Consider the example of two companies and the joint network of contacts between their employees. The code of conduct in each company may be different, depending on its internal organization and goals. Nevertheless, employees from one company may interact with employees from the other company, because they know each other personally or for instance when establishing a business transaction. By interacting with an acquaintance from the other company, each employee can acquire new working strategies, which can afterwards be imitated by their contacts inside each company. Another natural example is interaction between subpopulations of different cultural backgrounds.


\begin{figure}
\includegraphics[scale=0.55]{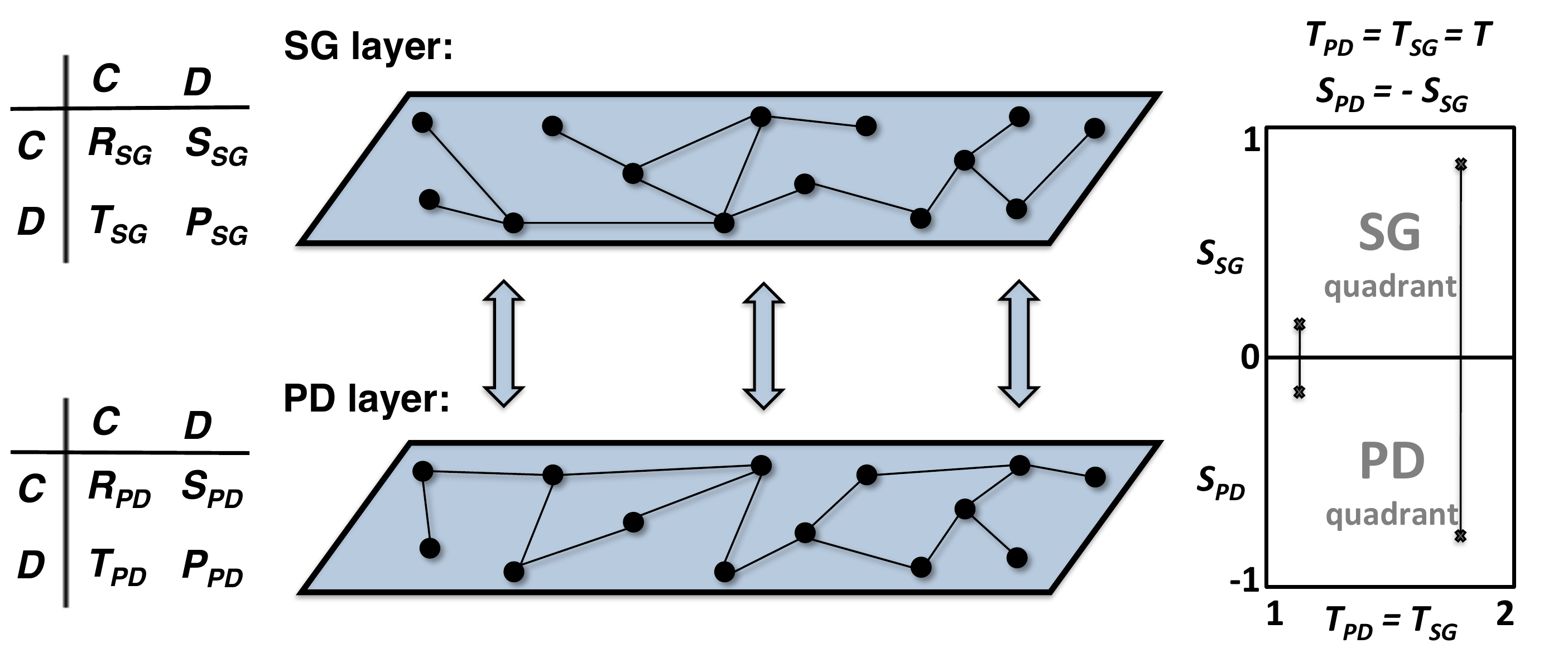}
\caption{\textbf{Scheme of the population structure.} The population is organized in two layers. Individuals establish {\it intralinks} with neighbors of the same layer, and {\it interlinks} with neighbors of the opposite layer. Depending on the layer in which they are located, individuals compute their payoff taking into account the PD payoff matrix or the SG payoff matrix, indicated on the left of the figure. On the right, we schematize the relation between the parameters of the two games, which we use in this paper. We assume that $T_{\ms{\text{PD}}}=T_{\ms{\text{SG}}}=T$, $S_{\ms{\text{SG}}}=S$ and $S_{\ms{\text{PD}}}=-S$, with $S \in \left[ 0,1 \right]$, $T \in \left[ 1,2 \right]$. \label{fig1}}
\end{figure}


Here we investigate how imitation and interaction between individuals playing different games influence the final cooperation levels. For that, we consider a population organized in two layers of equal size, in which one distinct game is played in each of the layers, as schematized in Fig.~\ref{fig1} (see also Methods for details). Nodes establish {\it intralinks} with neighbors of the same layer, and {\it interlinks} with neighbors of the opposite layer. In the following, we start by studying by computer simulations the case in which both layers are regular random graphs. Afterwards, we also consider the limit case of a well-mixed population of infinite size, also organized in two layers. We introduce bias in imitation: individuals imitate a neighbor from the same layer with probability $p$, or a neighbor of the opposite layer with probability $\left( 1 - p \right)$. We investigate the impact of varying $p$ on the final level of cooperation reached in each of the layers (see Methods).


\section{Results}

\subsection{Regular random layers}

\begin{figure}
\includegraphics[scale=0.5]{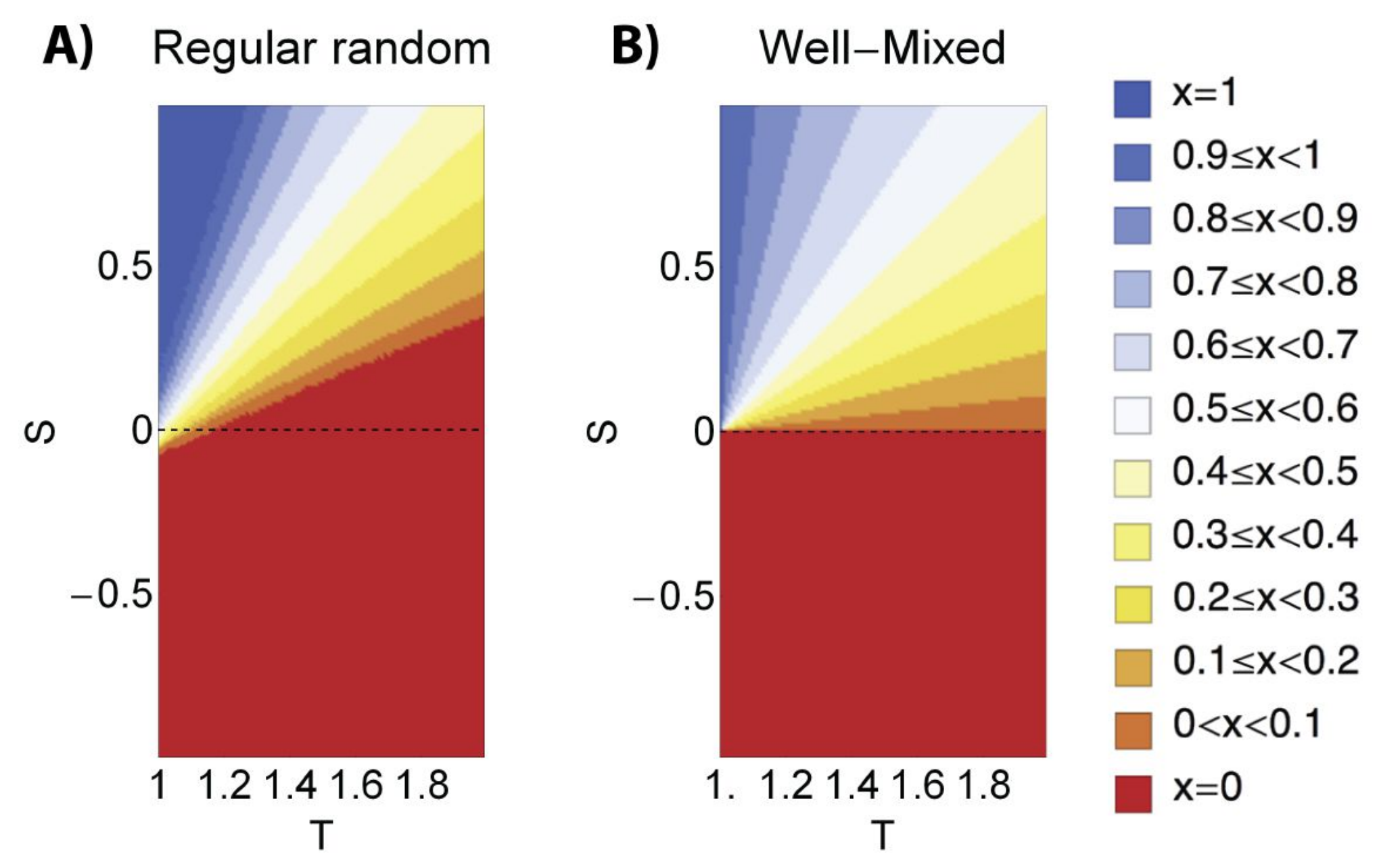}
\caption{\textbf{Final fraction of cooperators in single, isolated populations.} The contour plot of panel A) shows the final average fraction of cooperators for the PD and SG when these social dilemmas are played in an isolated regular random network of size $N = 10^{3}$ and node degree $k=4$. Panel B) shows the corresponding results for an infinite well-mixed population, solution of the replicator equation 
$\dot{x} = x \left( 1 - x \right) [ f_{\ms{\text{C}}} \left( x \right) - f_{\ms{\text{D}}} \left( x \right) ]$, 
where $f_{\ms{\text{C}}}(x)$ and $f_{\ms{\text{D}}}(x)$ stand for the mean fitness of cooperators and defectors respectively\cite{taylor1978evolutionary}. Red corresponds to full defection, blue corresponds to full cooperation. The parameter $\beta$ is set to $1.0$ for panel A). 
\label{traditionalresHoRandWM}
}
\end{figure}

For reference, we present in Fig.~\ref{traditionalresHoRandWM} the final fraction of cooperators for the PD and the SG played in isolated populations. Panel A) shows the results of computer simulations in regular random networks, while panel B) shows the stationary solutions of the replicator equation $\dot{x}= x \left( 1 - x \right) [ f_{\ms{\text{C}}} ( x ) - f_{\ms{\text{D}}} ( x ) ] $ for an infinite well-mixed population, that is a population in which every individual has the same probability of interacting with anyone else \cite{taylor1978evolutionary}. In the replicator equation, $x$ represents the fraction of cooperators in the population, $f_{\ms{\text{C}}}(x)$ and $f_{\ms{\text{D}}}(x)$ stand for the mean fitness of cooperators and defectors respectively (with $f_{\ms{\text{C}}} (x) = x R + (1-x)S$ and $f_{\ms{\text{D}}} = x T + (1-x)P$). The PD is the harshest social dilemma for cooperation, and therefore only a small region characterized by $S$ close to $0$ and $T$ close to $1$ is able to escape full defection, in regular random networks. The SG, on the other hand, is a coexistence game, in which the final fraction of cooperators $x$ in an infinite well-mixed population is given by 
$x=(P-S)/(R-S-T+P)$.


To evaluate the impact of an interdependence between the PD and the SG on the evolution of cooperation, we start by considering a population in which each of the layers is modeled by a regular random graph with nodes of degree $k=4$. This graph has no finite loops in the infinite size limit. We assume that each node establishes one interlink with a uniformly randomly chosen node from the opposite layer. That is, individuals engage in four interactions with neighbors of the same layer and one interaction with a neighbor from the other layer.

\begin{figure}
\includegraphics[scale=0.26]{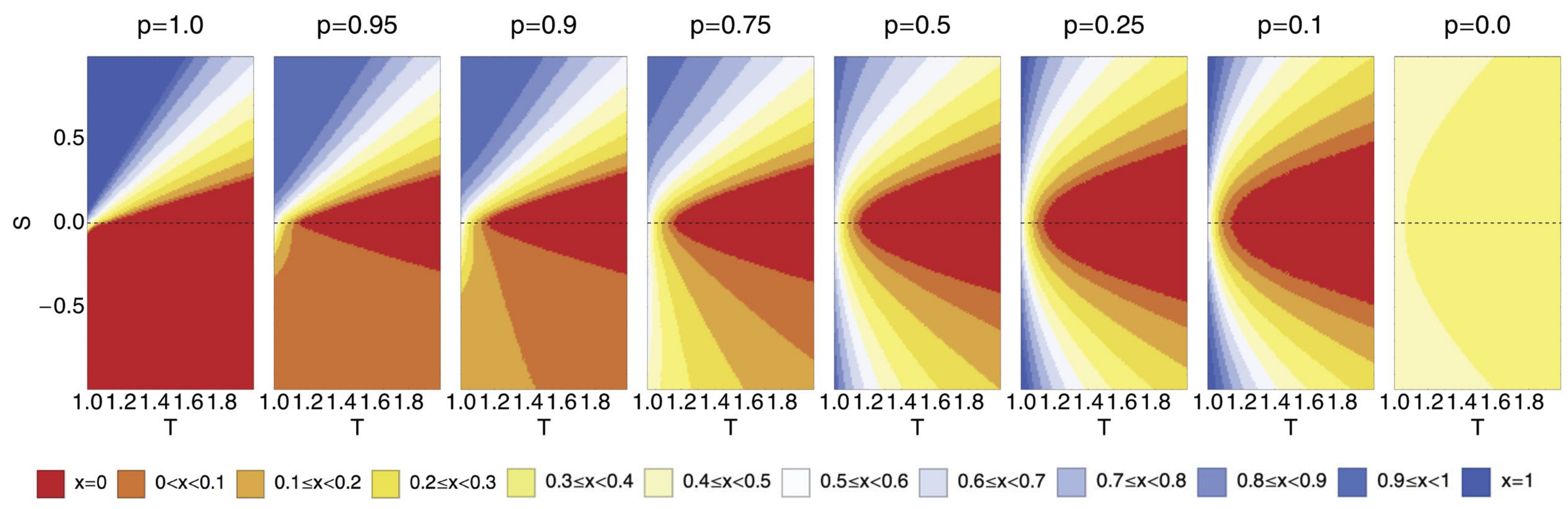}
\caption{\textbf{Final fraction of cooperators when layers are modelled as random regular graphs, for various values of probability $p$.} Contour plots show the final average fraction of cooperators on both the PD $\left( S < 0 \right)$ and SG $\left( S > 0 \right)$ layers as a function of the game parameters $T$ and $S$, starting from $50\%$ of cooperators and defectors randomly distributed in each of the layers. The dashed line separates the results for the SG layer (upper quadrants) from the results for the PD layer (lower quadrants). Red corresponds to full defection, blue corresponds to full cooperation ($x$ stands for the fraction of cooperators in a given layer). \label{fig3}}
\end{figure}

The contour plots in Fig.~\ref{fig3} show the result of our simulations, namely, the final fraction of cooperators in each of the layers as a function of the game parameters $T$ and $S$, for several biased imitation probabilities $p$ $\left( 0 \leq p \leq 1 \right)$. When $p=1$ individuals can only imitate neighbors from the same layer, and the results obtained are qualitatively similar to those known for isolated populations (see panel A) of Fig.~\ref{traditionalresHoRandWM}).


When $p<1$ strategy transfer between layers becomes possible. Comparing the results for the PD layer for $p=0.9$, $p=0.95$ and $p=1$, we reach an interesting conclusion: although the fraction of cooperators is still low, even a small probability of imitating a neighbor playing a distinct social dilemma (in this case, the SG) is sufficient to prevent full defection in the majority of the PD quadrant.

As $p$ further decreases, for a given pair of game parameters $T,S$ the fraction of cooperators in the stationary state becomes approximately equal in both layers (note that $S_{\ms{\text{SG}}} = S, S_{\ms{\text{PD}}} = -S$ with $S \in \left[ 0 , 1 \right]$). However, a careful inspection of the contour plots presented in Fig.~\ref{fig3} for $0.1 \leq p \leq 0.5$ shows that the fraction of cooperators in both layers is not exactly identical (i.e. the contours are not exactly symmetrical): for a fixed pair of $T$ and $S$ values, the PD layer tends to have a slightly lower fraction of cooperators.

\begin{figure}
\includegraphics[scale=0.65]{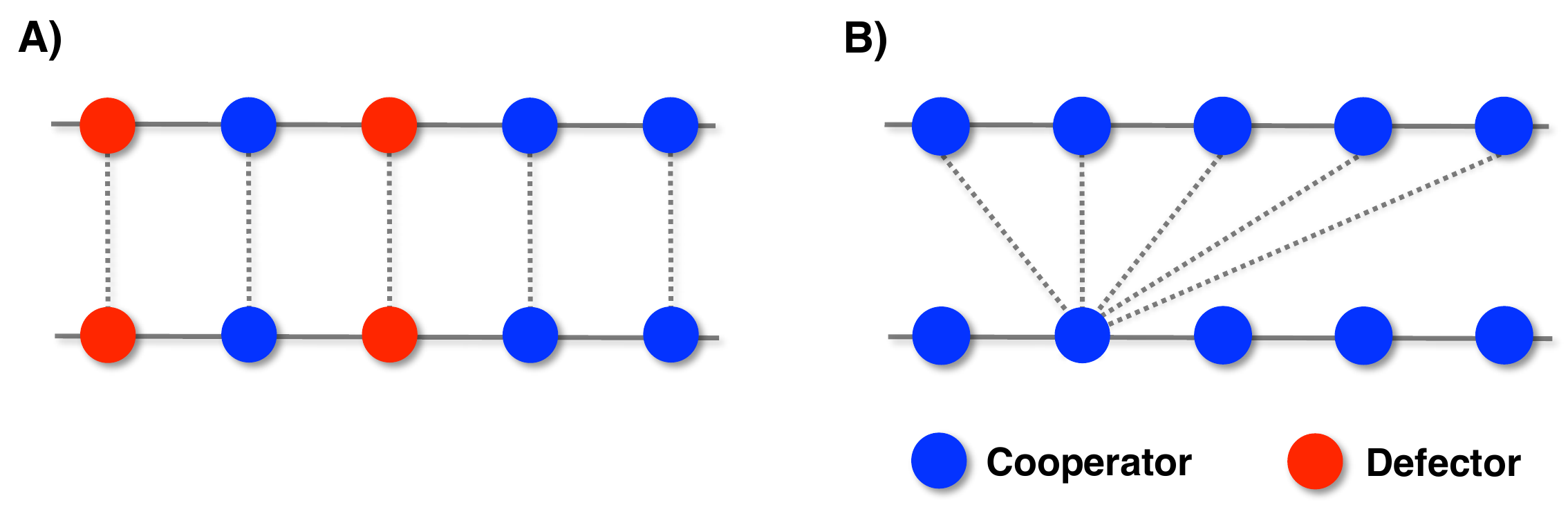}
\caption{\textbf{Absorbing states for $p=0$.} Each panel represents a snapshot of the two layers seen in profile. Panel A) schematizes one possible absorbing state when layers are modelled as regular random networks. Since each individual can only imitate his neighbor connected through the interlink, the population reaches an absorbing state organized as an arbitrary set of independent pairs (cooperator--cooperator and defector--defector) in which one member is in one layer and the other is in the other layer. Panel B) represents one of the two possible absorbing states in a well-mixed population. Since each node establishes interlinks to every node from the opposite layer, evolution ceases when both layers reach full cooperation or full defection.
}
\label{absorbingstates}
\end{figure}

Finally, the results for $p=0$ are surprisingly different from the ones for higher $p$. This contrast indicates the different structures of absorbing states at different values of $p$. When $p=1$ four absorbing states exist, and both layers reach homogeneous states: both reaching full defection (full cooperation), or one layer reaching full cooperation and the other full defection (and vice versa). When $0 < p < 1$, individuals can imitate neighbors from both layers, and therefore only the two absorbing states in which both layers reach the same homogeneous state are possible. In contrast, when $p=0$, each individual can only imitate his neighbor connected through the interlink. As a result, there is an infinite number of absorbing states. When all interlinks connect pairs of individuals with equal strategies, evolution stops, as schematized in panel A) of Fig.~\ref{absorbingstates}. Since strategies and interlinks are distributed randomly in the population, on average less than $N/2$ interlinks will actually connect individuals of different strategies, and the population will rapidly evolve to an absorbing state.


\begin{figure}
\includegraphics[scale=0.45]{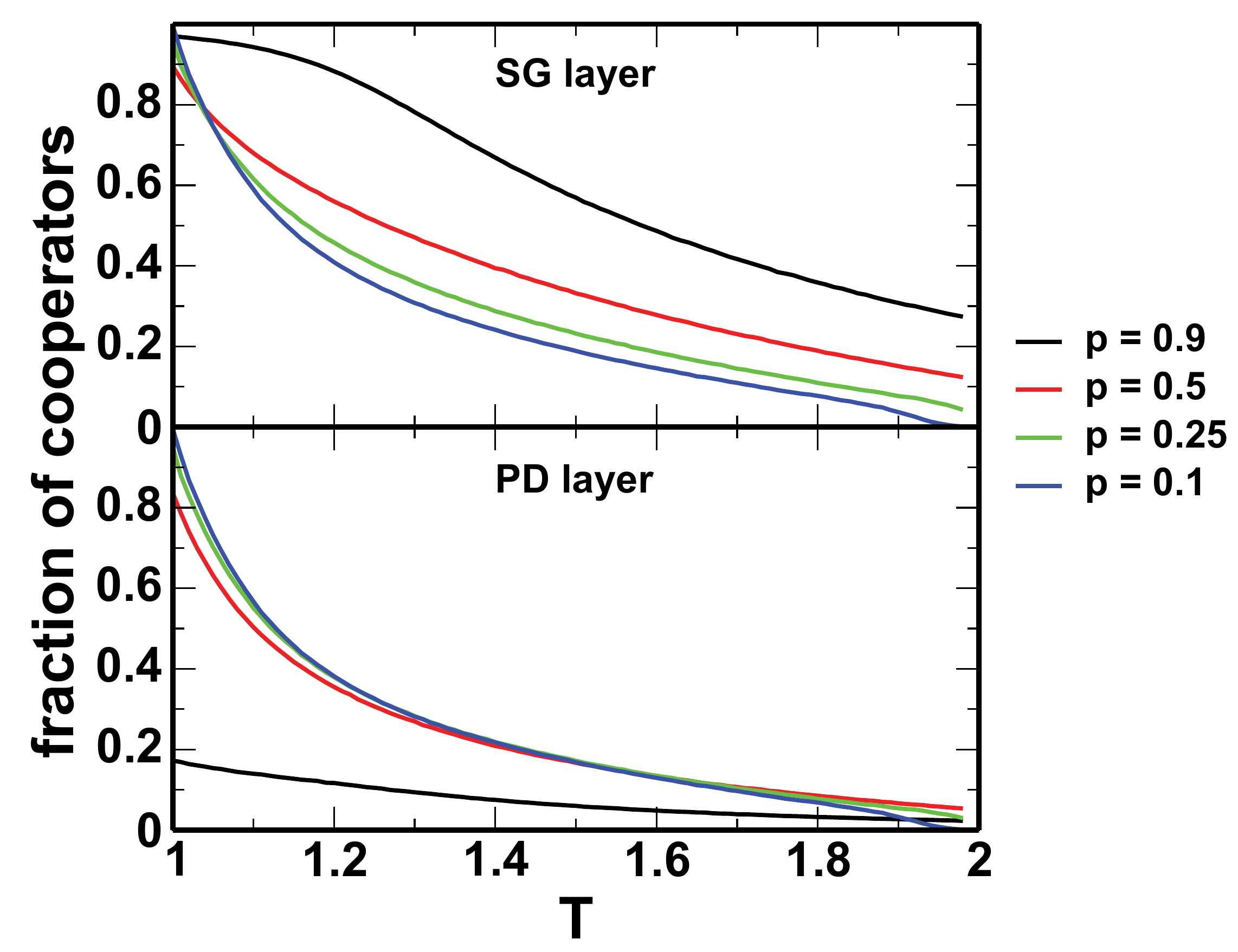}
\caption{\textbf{Final fraction of cooperators versus $T$ at fixed $S=0.5$, for both layers.} Results correspond to those presented in Fig.~\ref{fig3} when $S=0.5$. The fraction of cooperators for the SG layer is shown in the upper plot, while the corresponding curves for the PD layer are presented in the bottom plot. For simplicity, we include only some of the $p$ values studied in Fig.~\ref{fig3}.
 \label{selectedcontourHoRand}}
\end{figure}

Figure~\ref{selectedcontourHoRand} shows the cross section of the contour plots of Fig.~\ref{fig3} at fixed $S=0.5$ for several $p$ values. This illustrates the different behavior of the level of cooperation in each of the layers for decreasing $p$. Excluding the results for $p=0$ 
(which are a direct consequence of allowing exactly one interlink for each node) we observe that the fraction of cooperators monotonously decreases for decreasing $p$ in the SG layer, while in the PD layer it monotonously increases. As we will see in the following, this behavior will not be maintained when this population structure is replaced by a well-mixed population.


\subsection{Well-mixed population}

We now proceed to analyze the general case of a population modelled by a fully connected network organized in two layers. Each layer has $N/2$ nodes and $\frac12(N/2)(N/2 -1)$ intralinks. Contrary to the previous case of regular random layers, in which each node established one interlink, we now assume that each node establishes $N/2$ interlinks to the $N/2$ nodes of the opposite layer, in accordance with the structure of a fully connected network. Therefore, each node has $N-1$ neighbors: $N/2 -1$ in the same layer as itself, and $N/2$ in the opposite layer. $n_{\ms{\text{PD}}}$ and $n_{\ms{\text{SG}}}$ stand for the number of cooperators in the PD (SG) layer, respectively. In this case, mean-field theory is exact and the evolution of population can be expressed completely by $n_{\ms{\text{PD}}} \left( t \right)$ and $n_{\ms{\text{SG}}} \left( t \right)$.

We obtain the solutions for the stationary state of this model in the infinite population limit, $N \rightarrow \infty$. For that, we adopted the variables $x_{\ms{\text{PD}}} = n_{\ms{\text{PD}}}/(N/2)$ and $x_{\ms{\text{SG}}}=n_{\ms{\text{SG}}}/(N/2)$ $\left( x_{\ms{\text{PD}}} \in \left[ 0,1 \right]\text{ and }x_{\ms{\text{SG}}} \in \left[ 0,1 \right] \right)$. The mean fitness of individuals in each of the layers is expressed in terms of the game parameters and the fraction of cooperators and defectors. In the PD layer, the fitnesses of cooperators $\left( f_{\ms{\text{C,PD}}} \right)$ and defectors $\left( f_{\ms{\text{D,PD}}}  \right)$ are expressed as:
\begin{align}
f_{\ms{\text{C,PD}}} &=x_{\ms{\text{PD}}} - \left( 1 - x_{\ms{\text{PD}}} \right) S + x_{\ms{\text{SG}}} - \left( 1 - x_{\ms{\text{SG}}} \right) S,\\
f_{\ms{\text{D,PD}}} &= x_{\ms{\text{PD}}}T + x_{\ms{\text{SG}}} T.
\end{align}
Similarly, in the SG layer the fitnesses of cooperators $\left( f_{\ms{\text{C,SG}}} \right)$  and defectors $\left( f_{\ms{\text{D,SG}}} \right)$ are 
\begin{align}
f_{\ms{\text{C,SG}}} &= x_{\ms{\text{SG}}} + \left( 1 - x_{\ms{\text{SG}}} \right) S + x_{\ms{\text{PD}}} + \left( 1 - x_{\ms{\text{PD}}} \right) S,\\
f_{\ms{\text{D,SG}}} &= x_{\ms{\text{SG}}}T + x_{\ms{\text{PD}}}T.
\end{align}
We obtain the following rate equations for the evolution of the fraction of cooperators in the PD and SG layers:
\begin{eqnarray}
\dot{x}_{\ms{\text{PD}}} 
&=& 
\left( 1 - x_{\ms{\text{PD}}} \right)  \left[ p x_{\ms{\text{PD}}} F \left({f_{\ms{\text{C,PD}}} - f_{\ms{\text{D,PD}}}} \right) + \left( 1 - p \right) x_{\ms{\text{SG}}} F \left( f_{\ms{\text{C,SG}}} - f_{\ms{\text{D,PD}}} \right) \right]
\nonumber
\\[5pt]
&-& x_{\ms{\text{PD}}} \left[ p \left( 1 - x_{\text{PD}} \right) F \left( f_{\ms{\text{D,PD}}} - f_{\ms{\text{C,PD}}} \right) + \left( 1 - p \right) \left( 1 - x_{\ms{\text{SG}}} \right) F \left( f_{\ms{\text{D,SG}}} - f_{\ms{\text{C,PD}}} \right) \right]
,
\label{eq::gradPD}
\\[11pt]
\dot{x}_{\ms{\text{SG}}} 
&=& 
\left( 1 - x_{\ms{\text{SG}}} \right) \left[ p x_{\ms{\text{SG}}} F \left( f_{\ms{\text{C,SG}}} - f_{\ms{\text{D,SG}}} \right) + \left( 1 - p \right) x_{\ms{\text{PD}}} F \left( f_{\ms{\text{C,PD}}} - f_{\ms{\text{D,SG}}} \right) \right]
\nonumber
\\[5pt]
&-& 
x_{\ms{\text{SG}}} \left[ p \left( 1 - x_{\ms{\text{SG}}} \right) F \left( f_{\ms{\text{D,SG}}} - f_{\ms{\text{C,SG}}} \right) + \left( 1 - p \right) \left( 1 - x_{\ms{\text{PD}}} \right) F \left( f_{\ms{\text{D,PD}}} - f_{\ms{\text{C,SG}}} \right) \right]
.
\label{eq::gradSG}
\end{eqnarray}
For instance, in Eq.~(\ref{eq::gradPD}) the factor $\left( 1 - x_{\ms{\text{PD}}} \right)$ stands for the probability of selecting a defector from the PD layer. The second factor, in the square brackets, accounts for the two possible scenarios: with probability $p$ the defector can imitate a cooperator from the PD layer, while with complementary probability $\left( 1 - p \right)$ the defector can imitate a cooperator that belongs to the SG layer. The interpretation is analogous for the second line of this equation, as well as for Eq. (\ref{eq::gradSG}). The stationary state solution corresponds to that obtained from Eqs. (\ref{eq::gradPD}) and (\ref{eq::gradSG}) when $\dot{x}_{\ms{\text{PD}}} = 0$ and $\dot{x}_{\ms{\text{SG}}}=0$.


\begin{figure}
\includegraphics[scale=0.25]{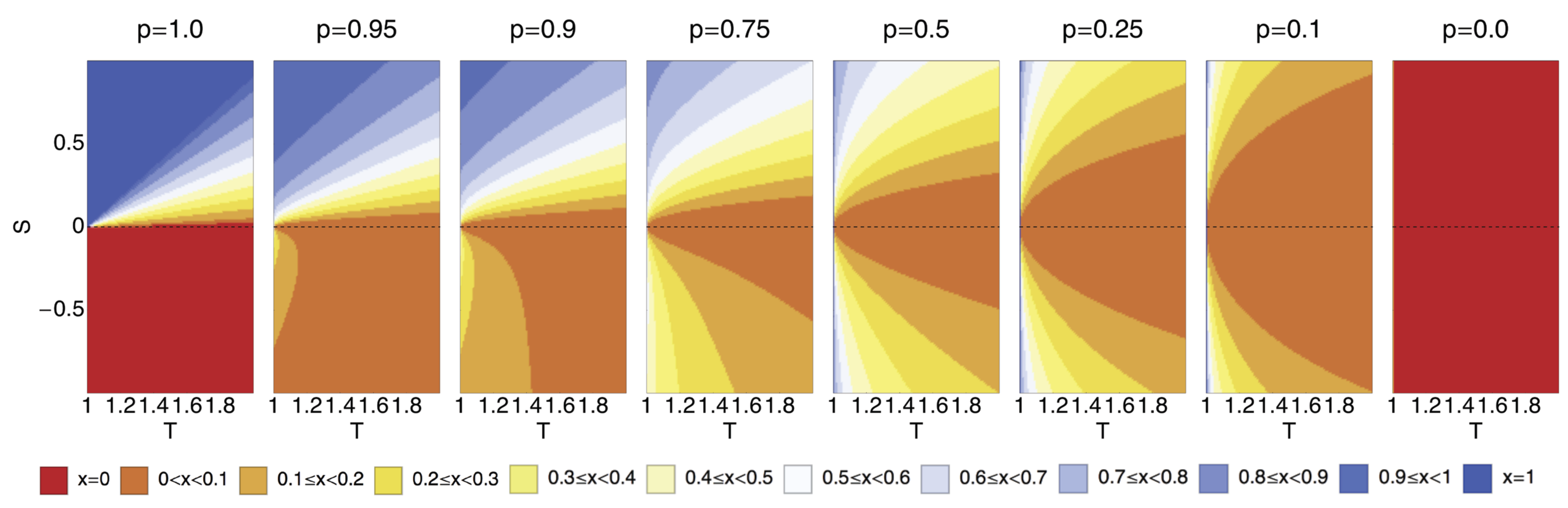}
\caption{\textbf{Final fraction of cooperators on each layer of the well-mixed population, for various values of probability $p$.} Contour plots show the final fraction of cooperators on both the PD $\left( S<0\right)$ and SG $\left( S>0\right)$ layers (solution of Eqs. (\ref{eq::gradPD}) and (\ref{eq::gradSG})) as a function of the game parameters $T$ and $S$. The dashed line separates the results for the SG layer (upper quadrants) from the results for the PD layer (lower quadrants). Red corresponds to full defection, blue corresponds to full cooperation ($x$ stands for the fraction of cooperators in a given layer). Parameters: $\beta = 1.0$.\label{fig2}}
\end{figure}

\begin{figure}
\includegraphics[scale=0.45]{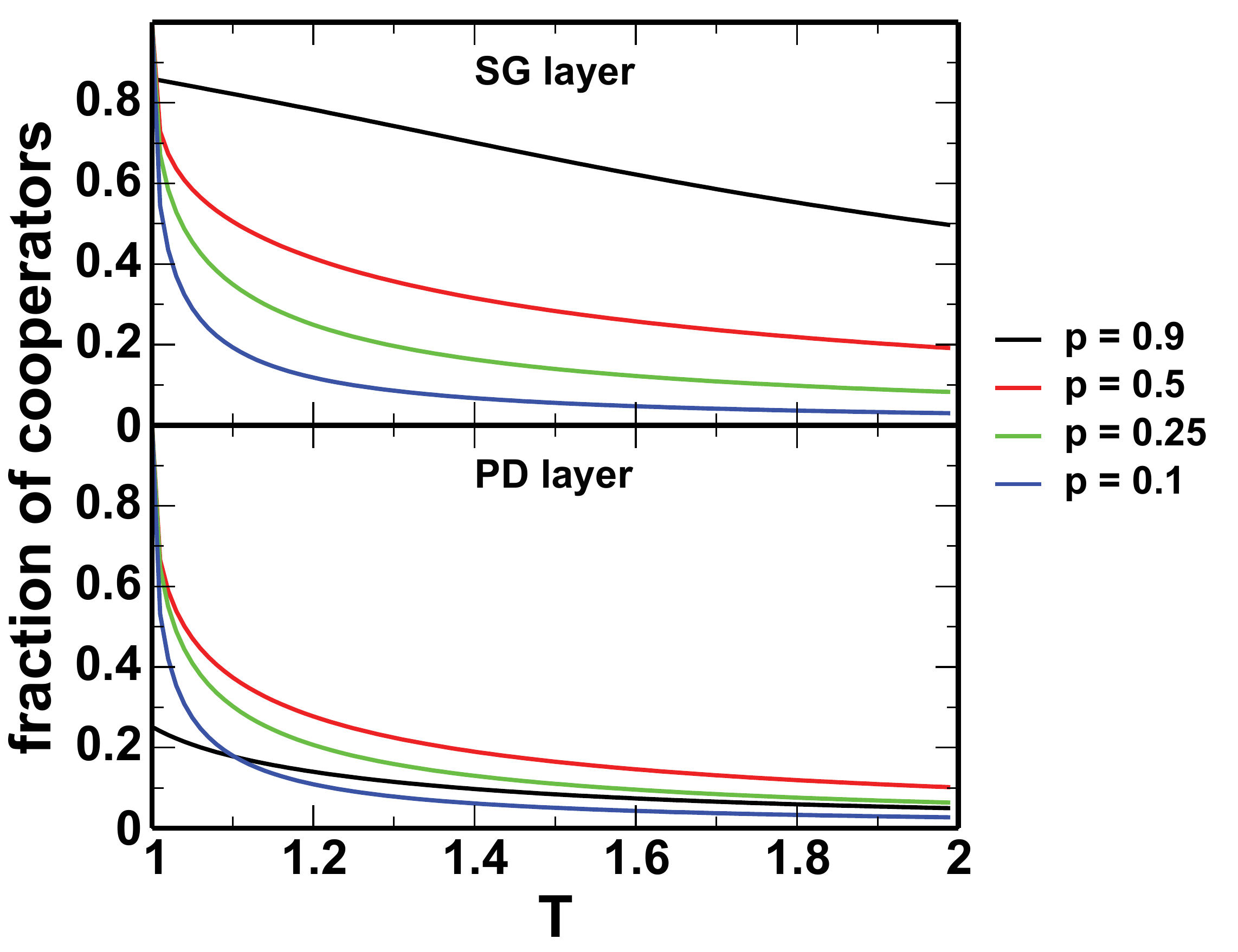}
\caption{
\textbf{Final fraction of cooperators versus $T$ at fixed $S=0.5$ for both layers of a fully connected network.} Results correspond to those presented in Fig.~\ref{fig2} when $S= 0.5$. The fraction of cooperators for the SG layer is shown in the upper plot, while the corresponding curves for the PD layer are shown in the bottom plot. 
}
\label{selectedcontourWM}
\end{figure}

Figure~\ref{fig2} shows the final stationary states of this mean-field model for different values of probability $p$, provided by the numerical solutions of Eqs. (\ref{eq::gradPD}) and (\ref{eq::gradSG}) for $\beta = 1.0$. These plots actually explain our simulations in the previous section for coupled regular random networks (see Fig.~\ref{fig3}). One can clearly see that Figs.~\ref{fig3} and~\ref{fig2} show qualitative similarities, with only one dramatic exception, $p=0$, which will be discussed below. In particular, in the case of $p=1$ both Figs.~\ref{fig3} and~\ref{fig2}, as well as Figs.~\ref{selectedcontourHoRand} and~\ref{selectedcontourWM}, are close to the corresponding results for isolated populations (random regular and well-mixed, respectively, compare with Fig.~\ref{traditionalresHoRandWM}). 

One can see from Fig.~\ref{fig2} that a small probability of imitating a neighbor from the SG layer is beneficial for the levels of cooperation reached in the PD layer. The final fraction of cooperators in each of the layers becomes approximately equal as $p$ decreases, yielding almost symmetrical contours for $0.1 \leq p \leq 0.5$. However, the fraction of cooperators is in general slightly lower in the PD layer than in the SG layer for a fixed pair of game parameters $T$ and $S$.

Note some difference between Figs.~\ref{fig3} and~\ref{fig2} at sufficiently low, still non-zero, $p$. With diminishing $p$, the region of full defection in coupled regular random populations (Fig.~\ref{fig3}) approaches some limit. In contrast, for decreasing $p$, coupled well-mixed populations (Fig.~\ref{fig2}) do not show any area with full defection, 
but the region with high level of defection becomes wider and wider as $p$ decreases. 
Finally, at $p=0$, the full defection state is realized in the entire $T-S$ plane. This difference is also observed between Figs.~\ref{selectedcontourHoRand} and~\ref{selectedcontourWM}. Especially it manifests itself in two different behaviors of fractions of cooperators in the PD layer in Figs.~\ref{selectedcontourHoRand} and~\ref{selectedcontourWM}, monotonous with $p$ and non-monotonous, respectively.

At $p=0$ the difference between the coupled regular random networks and well-mixed population is dramatic. While for the coupled well-mixed populations, the final state is full defection for the whole $T-S$ plane (Fig.~\ref{fig2}), for regular random networks we observed coexistence of defection and cooperation for the whole $T-S$ plane. The reason for this controversy is the fundamental difference between the structures of the absorbing states in these two systems. Clearly, in the mean-field system (well-mixed) there are two absorbing states: full cooperation and full defection. For $p=0$, the full defection scenario is realized. In contrast, in the coupled regular random populations, at $p=0$ the system has an infinite number of absorbing states. We already explained that this state is organized as an arbitrary set of independent pairs (cooperator-cooperator and defector-defector) in which one member is in one layer and the other is in the other layer. The final fraction of cooperators and defectors in the population for a given point on the $T-S$ plane is determined by a specific initial condition. 

It is interesting to relate our findings with observations in other works \cite{wang2013optimal,wang2012evolution} which studied the role of biased fitness (utility) on the level of cooperation in interdependent structured populations. Although these works explored a bias and network architectures very different from our study, they reported the existence of an optimal control parameter value for cooperation, similarly to what we notice for a well-mixed population. 

We can also relate our model to the multi-network model \cite{ohtsuki2007breaking,ohtsuki2007evolutionary}, where links between nodes are of two sorts: the interaction links through which individuals play and the imitation links through which individuals revise their strategies. All links in our model are used both for interaction and imitation. The principal difference is that interaction occurs with probability $1$ for every link, while imitation takes place with probability $p$ along intralinks and the complementary probability $1-p$ along interlinks. 


\section{Discussion}

We explored the evolution of cooperation in two games---the Prisoner's Dilemma and the Snowdrift Game---played within an interdependent network. We demonstrated that as the probability to imitate neighbors from the opposite layer increases, 
the final level of cooperation in each of the layers behaves differently. A small probability $1-p$ is always detrimental for the level of cooperation in the SG layer, regardless of the population structure (well-mixed or when layers are modelled as regular random networks). In the case of a well-mixed population we observed an intermediate optimal value of $p$ for which the fraction of cooperators in the steady state is maximal in the PD quadrant. 
On the other hand, when layers are regular random graphs, the level of cooperation in the PD layer registers a significant increase from full defection as $p$ decreases from $1$, as is shown in Fig.~\ref{fig3}. We have also shown that different arrangements of interconnections between layers result in fundamentally different structures of absorbing states for $p=0$, that is the limit case in which individuals can only imitate neighbors through the interlinks. 

The constraint between the game parameters of the PD and the SG implemented in this work is one of possible options. Based on our mean-field approach we have verified that our conclusions remain qualitatively valid for other relationships between game parameters \cite{xxx}. In this work we studied a representative case of $\beta=1$. In the Supplementary Information we consider other values of $\beta$.  

The Prisoner's Dilemma and the Snowdrift Game are two of the possible two-person games. Future research directions include establishing an interdependence between other two- and $N$-person games, as well as extending the biased imitation process here reported to the case in which layers are strongly heterogeneous, such as scale-free networks \cite{barabasi1999emergence}. 


\section{Methods} 

\textbf{Evolutionary games.} 
We implement standard two-person games. 
Each individual can act as an unconditional cooperator ({\it \textbf{C}}) or as an unconditional defector ({\it \textbf{D}}). The possible payoffs resulting from the interaction between two players are resumed in the payoff matrix
\begin{equation}
\label{eq::generalpayoffmatrix}
\bordermatrix{ \text{ }&\mathbi{C}&\mathbi{D}\cr
  \mathbi{C}&R & S \cr
  \mathbi{D}&T & P \\}
\end{equation}
in which the entries represent the payoff earned by the row player. When both individuals opt for cooperation (defection), they both receive a Reward, $R$ (Punishment, $P$). When a cooperator meets a defector, the cooperative player received the Sucker's payoff ($S$), while the defective player receives the Temptation ($T$) for defecting.

The relative ordering of the game parameters $R$, $S$, $T$ and $P$ defines four different two-person games \cite{kollock1998social}. In this study, we consider the Prisoner's Dilemma (PD) (characterized by $T>R>P>S$) \cite{rapoport1965prisoner} and the Snowdrift Game (SG) (characterized by $T>R>S>P$) \cite{sugden1986economics}. We fix $R=1$, $P=0$ as is conventional \cite{nowak1992evolutionary}.

\textbf{Population structure.} 
We consider a population organized in two layers (schematized in Fig.~\ref{fig1}), in which each node represents a different individual. 
In each layer all individuals play the same game, the PD in one layer and the SG in the other. We refer to links between nodes of opposite layers as {\it interlinks}, and links between nodes of the same layer as {\it intralinks}. In the absence of interlinks, each layer stands as an isolated population in which one single game is played. When layers are modelled as regular random graphs, we assume that each node establishes one interlink to a randomly chosen node from the opposite layer. On the other hand, when studying a two-layer fully connected network, we assume that each node establishes interlinks to each of the nodes in the opposite layer.

\textbf{Relation between game parameters.} 
We assume that individuals in the PD layer adopt the game parameters $R_{\ms{\text{PD}}}$, $S_{\ms{\text{PD}}}$, $T_{\ms{\text{PD}}}$ and $P_{\ms{\text{PD}}}$ (with $T_{\ms{\text{PD}}} > R_{\ms{\text{PD}}} > P_{\ms{\text{PD}}} > S_{\ms{\text{PD}}}$). Similarly, individuals in the SG layer consider $R_{\ms{\text{SG}}}$, $S_{\ms{\text{SG}}}$, $T_{\ms{\text{SG}}}$ and $P_{\ms{\text{SG}}}$ (with $T_{\ms{\text{SG}}} > R_{\ms{\text{SG}}} > S_{\ms{\text{SG}}} > P_{\ms{\text{SG}}}$). The game parameters for the PD and the SG can be related in several ways. We opt for $R_{\ms{\text{PD}}} = R_{\ms{\text{SG}}} = 1$, $P_{\ms{\text{PD}}}=P_{\ms{\text{SG}}}=0$ and $T_{\ms{\text{PD}}}=T_{\ms{\text{SG}}}=T$ with $T \in \left[ 1,2 \right]$. Furthermore, we constrained $S_{\ms{\text{SG}}} = S$ and $S_{\ms{\text{PD}}} = -S$, with $S \in \left[ 0,1 \right]$. The resulting payoff matrix for the SG is 
\begin{equation}
\label{eq::paymatrixSG}
\bordermatrix{ \text{ }&\mathbi{C}&\mathbi{D}\cr
  \mathbi{C}&1 & S \cr
  \mathbi{D}&T & 0 \\}
\end{equation}
while the payoff matrix for the PD layer is given by
\begin{equation}
\label{eq::paymatrixPD}
\bordermatrix{ \text{ }&\mathbi{C}&\mathbi{D}\cr
  \mathbi{C}&1 & -S \cr
  \mathbi{D}&T & 0 \\}
  .
\end{equation}
The difference between matrix (\ref{eq::paymatrixSG}) for the SG layer and matrix (\ref{eq::paymatrixPD}) for the PD layer is in the payoff obtained by a cooperator when interacting with a defector, the sucker's payoff. When interacting with a defector, a cooperator located in the SG layer earns $S$, while a cooperator located in the PD layer earns $-S$. We stress that this is only one of possible relations between the parameters of the two games.

Individuals play the same game with all neighbors, including the neighbor(s) from the other layer. An individual's fitness represents a measure of his success, and corresponds to the payoff accumulated over all interactions in which he engages in one time-step. Through the interlinks, individuals obtain their payoff by taking into account the strategy of their co-player and the payoff matrix used in their own layer.

\textbf{Update dynamics.} Individuals revise their strategies through a {\it biased imitation process}: with probability $p$ they imitate a neighbor from the same layer, while with probability $\left( 1-p \right)$ individuals imitate a neighbor from the opposite layer. Update is asynchronous: at each time-step we randomly choose one of the layers, and from it one randomly chosen individual, A, to revise his strategy. Then we choose with probability $p$ a neighbor from the same layer, and with complimentary probability $\left( 1 - p \right)$, a neighbor from the opposite layer. If A and his chosen neighbor, B, have different strategies, A imitates the strategy of B with a probability $F \left( f_{\ms{\text{B}}} - f_{\ms{\text{A}}} \right)$ proportional to their fitness difference and given by the Fermi distribution \cite{traulsen2007stochastic,traulsen2007pairwise},
\begin{equation}
F \left( f_{\ms{\text{B}}} - f_{\ms{\text{A}}} \right)=\frac{1}{1+e^{- \beta \left( f_{\text{B}} - f_{\text{A}} \right)}}
.
\end{equation}
$f_{\text{A}}$ $\left( f_{\text{B}} \right)$ stands for the fitness of A (B), and $\beta$ represents the intensity of selection, which regulates the accuracy of the imitation process. In the limit $\beta \rightarrow 0$ evolution proceeds by random drift, while $\beta \rightarrow \infty$ corresponds to the deterministic limit of pure copying dynamics: A imitates B iff $f_{\text{B}} > f_{\text{A}}$, with probability 1. We consider the respresentative case of $\beta=1.0$.

\textbf{Simulations.} The results of Figs.~\ref{fig3} and ~\ref{selectedcontourHoRand} were obtained for a population of total size $N=2000$, in which both layers had an equal size, $N/2 = 1000$. Each point in the contour plots corresponds to an average over $10^{3}$ trials. In each trial, we start by randomly selecting one over $10^{2}$ realizations of a regular random network to model each layer, and establish $N/2$ interlinks between the layers. Each interlink is established by uniformly randomly choosing two end nodes, one from each layer, allowing each node to establish exactly one interlink. Each trial starts by randomly distributing $50\%$ of cooperators and defectors in each of the layers. We allow the population to evolve for $10^{4}$ generations (where 1 generation corresponds to $N/2$ strategy revisions), after which we average the fraction of cooperators over the next $10^{3}$ generations.


\begin{thebibliography}{10}
\expandafter\ifx\csname url\endcsname\relax
  \def\url#1{\texttt{#1}}\fi
\expandafter\ifx\csname urlprefix\endcsname\relax\def\urlprefix{URL }\fi
\providecommand{\bibinfo}[2]{#2}
\providecommand{\eprint}[2][]{\url{#2}}

\bibitem{buldyrev2010catastrophic}
\bibinfo{author}{Buldyrev, S.~V.}, \bibinfo{author}{Parshani, R.},
  \bibinfo{author}{Paul, G.}, \bibinfo{author}{Stanley, H.~E.} \&
  \bibinfo{author}{Havlin, S.}
\newblock \bibinfo{title}{Catastrophic cascade of failures in interdependent
  networks}.
\newblock \emph{\bibinfo{journal}{Nature}} \textbf{\bibinfo{volume}{464}},
  \bibinfo{pages}{1025--1028} (\bibinfo{year}{2010}).

\bibitem{PhysRevLett.111.058701}
\bibinfo{author}{Nicosia, V.}, \bibinfo{author}{Bianconi, G.},
  \bibinfo{author}{Latora, V.} \& \bibinfo{author}{Barthelemy, M.}
\newblock \bibinfo{title}{Growing multiplex networks}.
\newblock \emph{\bibinfo{journal}{Phys. Rev. Lett.}}
  \textbf{\bibinfo{volume}{111}}, \bibinfo{pages}{058701}
  (\bibinfo{year}{2013}).

\bibitem{cozzo2013contact}
\bibinfo{author}{Cozzo, E.}, \bibinfo{author}{Banos, R.~A.},
  \bibinfo{author}{Meloni, S.} \& \bibinfo{author}{Moreno, Y.}
\newblock \bibinfo{title}{Contact-based social contagion in multiplex
  networks}.
\newblock \emph{\bibinfo{journal}{Phys. Rev. E}} \textbf{\bibinfo{volume}{88}},
  \bibinfo{pages}{050801} (\bibinfo{year}{2013}).

\bibitem{helbing2013globally}
\bibinfo{author}{Helbing, D.}
\newblock \bibinfo{title}{Globally networked risks and how to respond}.
\newblock \emph{\bibinfo{journal}{Nature}} \textbf{\bibinfo{volume}{497}},
  \bibinfo{pages}{51--59} (\bibinfo{year}{2013}).

\bibitem{baxter2012avalanche}
\bibinfo{author}{Baxter, G.~J.}, \bibinfo{author}{Dorogovtsev, S.~N.},
  \bibinfo{author}{Goltsev, A.~V.} \& \bibinfo{author}{Mendes, J. F.~F.}
\newblock \bibinfo{title}{Avalanche collapse of interdependent networks}.
\newblock \emph{\bibinfo{journal}{Phys. Rev. Lett.}}
  \textbf{\bibinfo{volume}{109}}, \bibinfo{pages}{248701}
  (\bibinfo{year}{2012}).

\bibitem{parshani2010interdependent}
\bibinfo{author}{Parshani, R.}, \bibinfo{author}{Buldyrev, S.~V.} \&
  \bibinfo{author}{Havlin, S.}
\newblock \bibinfo{title}{Interdependent networks: reducing the coupling
  strength leads to a change from a first to second order percolation
  transition}.
\newblock \emph{\bibinfo{journal}{Phys. Rev. Lett.}}
  \textbf{\bibinfo{volume}{105}}, \bibinfo{pages}{048701}
  (\bibinfo{year}{2010}).

\bibitem{son2012percolation}
\bibinfo{author}{Son, S.-W.}, \bibinfo{author}{Bizhani, G.},
  \bibinfo{author}{Christensen, C.}, \bibinfo{author}{Grassberger, P.} \&
  \bibinfo{author}{Paczuski, M.}
\newblock \bibinfo{title}{Percolation theory on interdependent networks based
  on epidemic spreading}.
\newblock \emph{\bibinfo{journal}{EPL}} \textbf{\bibinfo{volume}{97}},
  \bibinfo{pages}{16006} (\bibinfo{year}{2012}).

\bibitem{gao2011robustness}
\bibinfo{author}{Gao, J.}, \bibinfo{author}{Buldyrev, S.~V.},
  \bibinfo{author}{Havlin, S.} \& \bibinfo{author}{Stanley, H.~E.}
\newblock \bibinfo{title}{Robustness of a network of networks}.
\newblock \emph{\bibinfo{journal}{Phys. Rev. Lett.}}
  \textbf{\bibinfo{volume}{107}}, \bibinfo{pages}{195701}
  (\bibinfo{year}{2011}).

\bibitem{zhou2012assortativity}
\bibinfo{author}{Zhou, D.}, \bibinfo{author}{Stanley, H.~E.},
  \bibinfo{author}{D'Agostino, G.} \& \bibinfo{author}{Scala, A.}
\newblock \bibinfo{title}{Assortativity decreases the robustness of
  interdependent networks}.
\newblock \emph{\bibinfo{journal}{Phys. Rev. E}} \textbf{\bibinfo{volume}{86}},
  \bibinfo{pages}{066103} (\bibinfo{year}{2012}).

\bibitem{li2012cascading}
\bibinfo{author}{Li, W.}, \bibinfo{author}{Bashan, A.},
  \bibinfo{author}{Buldyrev, S.~V.}, \bibinfo{author}{Stanley, H.~E.} \&
  \bibinfo{author}{Havlin, S.}
\newblock \bibinfo{title}{Cascading failures in interdependent lattice
  networks: The critical role of the length of dependency links}.
\newblock \emph{\bibinfo{journal}{Phys. Rev. Lett.}}
  \textbf{\bibinfo{volume}{108}}, \bibinfo{pages}{228702}
  (\bibinfo{year}{2012}).

\bibitem{huang2011robustness}
\bibinfo{author}{Huang, X.}, \bibinfo{author}{Gao, J.},
  \bibinfo{author}{Buldyrev, S.~V.}, \bibinfo{author}{Havlin, S.} \&
  \bibinfo{author}{Stanley, H.~E.}
\newblock \bibinfo{title}{Robustness of interdependent networks under targeted
  attack}.
\newblock \emph{\bibinfo{journal}{Phys. Rev. E}} \textbf{\bibinfo{volume}{83}},
  \bibinfo{pages}{065101} (\bibinfo{year}{2011}).

\bibitem{vespignani2010complex}
\bibinfo{author}{Vespignani, A.}
\newblock \bibinfo{title}{Complex networks: The fragility of interdependency}.
\newblock \emph{\bibinfo{journal}{Nature}} \textbf{\bibinfo{volume}{464}},
  \bibinfo{pages}{984--985} (\bibinfo{year}{2010}).

\bibitem{radicchi2013abrupt}
\bibinfo{author}{Radicchi, F.} \& \bibinfo{author}{Arenas, A.}
\newblock \bibinfo{title}{Abrupt transition in the structural formation of
  interconnected networks}.
\newblock \emph{\bibinfo{journal}{Nat. Phys.}} \textbf{\bibinfo{volume}{9}},
  \bibinfo{pages}{717--720} (\bibinfo{year}{2013}).

\bibitem{poljanvsek2012seismic}
\bibinfo{author}{Poljan{\v{s}}ek, K.}, \bibinfo{author}{Bono, F.} \&
  \bibinfo{author}{Guti{\'e}rrez, E.}
\newblock \bibinfo{title}{Seismic risk assessment of interdependent critical
  infrastructure systems: The case of european gas and electricity networks}.
\newblock \emph{\bibinfo{journal}{Earthquake Eng. Struct. Dynam.}}
  \textbf{\bibinfo{volume}{41}}, \bibinfo{pages}{61--79}
  (\bibinfo{year}{2012}).

\bibitem{duenas2007seismic}
\bibinfo{author}{Due{\~n}as-Osorio, L.}, \bibinfo{author}{Craig, J.~I.} \&
  \bibinfo{author}{Goodno, B.~J.}
\newblock \bibinfo{title}{Seismic response of critical interdependent
  networks}.
\newblock \emph{\bibinfo{journal}{Earthquake Eng. Struct. Dynam.}}
  \textbf{\bibinfo{volume}{36}}, \bibinfo{pages}{285--306}
  (\bibinfo{year}{2007}).

\bibitem{csermely2012appearance}
\bibinfo{author}{Csermely, P.}
\newblock \bibinfo{title}{The appearance and promotion of creativity at various
  levels of interdependent networks}.
\newblock \emph{\bibinfo{journal}{Talent Development Excellence}}
  \textbf{\bibinfo{volume}{5}}, \bibinfo{pages}{115--123}
  (\bibinfo{year}{2013}).

\bibitem{halu2013connect}
\bibinfo{author}{Halu, A.}, \bibinfo{author}{Zhao, K.},
  \bibinfo{author}{Baronchelli, A.} \& \bibinfo{author}{Bianconi, G.}
\newblock \bibinfo{title}{Connect and win: The role of social networks in
  political elections}.
\newblock \emph{\bibinfo{journal}{EPL}} \textbf{\bibinfo{volume}{102}},
  \bibinfo{pages}{16002} (\bibinfo{year}{2013}).

\bibitem{dorogovtsev2002evolution}
\bibinfo{author}{Dorogovtsev, S.~N.} \& \bibinfo{author}{Mendes, J.~F.}
\newblock \bibinfo{title}{Evolution of networks}.
\newblock \emph{\bibinfo{journal}{Adv. Phys.}} \textbf{\bibinfo{volume}{51}},
  \bibinfo{pages}{1079--1187} (\bibinfo{year}{2002}).

\bibitem{dorogovtsev2003evolution}
\bibinfo{author}{Dorogovtsev, S.~N.} \& \bibinfo{author}{Mendes, J.~F.}
\newblock \emph{\bibinfo{title}{Evolution of Networks: From Biological Nets to
  the Internet and WWW}} (\bibinfo{publisher}{Oxford University Press},
  \bibinfo{year}{2003}).

\bibitem{albert2002statistical}
\bibinfo{author}{Albert, R.} \& \bibinfo{author}{Barab{\'a}si, A.-L.}
\newblock \bibinfo{title}{Statistical mechanics of complex networks}.
\newblock \emph{\bibinfo{journal}{Rev. Mod. Phys.}}
  \textbf{\bibinfo{volume}{74}}, \bibinfo{pages}{47} (\bibinfo{year}{2002}).

\bibitem{sigmund2010calculus}
\bibinfo{author}{Sigmund, K.}
\newblock \emph{\bibinfo{title}{The Calculus of Selfishness}}
  (\bibinfo{publisher}{Princeton University Press}, \bibinfo{year}{2010}).

\bibitem{axelrod1981evolution}
\bibinfo{author}{Axelrod, R.} \& \bibinfo{author}{Hamilton, W.~D.}
\newblock \bibinfo{title}{The evolution of cooperation}.
\newblock \emph{\bibinfo{journal}{Science}} \textbf{\bibinfo{volume}{211}},
  \bibinfo{pages}{1390--1396} (\bibinfo{year}{1981}).

\bibitem{rapoport1965prisoner}
\bibinfo{author}{Rapoport, A.}
\newblock \emph{\bibinfo{title}{Prisoner's Dilemma: A Study in Conflict and
  Cooperation}} (\bibinfo{publisher}{University of Michigan Press},
  \bibinfo{year}{1965}).

\bibitem{sugden1986economics}
\bibinfo{author}{Sugden, R.}
\newblock \emph{\bibinfo{title}{The Economics of Rights, Co-operation and
  Welfare}} (\bibinfo{publisher}{Basil Blackwell Oxford},
  \bibinfo{year}{1986}).

\bibitem{skyrms2004stag}
\bibinfo{author}{Skyrms, B.}
\newblock \emph{\bibinfo{title}{The Stag Hunt and the Evolution of Social
  Structure}} (\bibinfo{publisher}{Cambridge University Press},
  \bibinfo{year}{2004}).

\bibitem{nowak1992evolutionary}
\bibinfo{author}{Nowak, M.~A.} \& \bibinfo{author}{May, R.~M.}
\newblock \bibinfo{title}{Evolutionary games and spatial chaos}.
\newblock \emph{\bibinfo{journal}{Nature}} \textbf{\bibinfo{volume}{359}},
  \bibinfo{pages}{826--829} (\bibinfo{year}{1992}).

\bibitem{nowak2006five}
\bibinfo{author}{Nowak, M.~A.}
\newblock \bibinfo{title}{Five rules for the evolution of cooperation}.
\newblock \emph{\bibinfo{journal}{Science}} \textbf{\bibinfo{volume}{314}},
  \bibinfo{pages}{1560--1563} (\bibinfo{year}{2006}).

\bibitem{santos2005scale}
\bibinfo{author}{Santos, F.~C.} \& \bibinfo{author}{Pacheco, J.~M.}
\newblock \bibinfo{title}{Scale-free networks provide a unifying framework for
  the emergence of cooperation}.
\newblock \emph{\bibinfo{journal}{Phys. Rev. Lett.}}
  \textbf{\bibinfo{volume}{95}}, \bibinfo{pages}{098104}
  (\bibinfo{year}{2005}).

\bibitem{santos2008social}
\bibinfo{author}{Santos, F.~C.}, \bibinfo{author}{Santos, M.~D.} \&
  \bibinfo{author}{Pacheco, J.~M.}
\newblock \bibinfo{title}{Social diversity promotes the emergence of
  cooperation in public goods games}.
\newblock \emph{\bibinfo{journal}{Nature}} \textbf{\bibinfo{volume}{454}},
  \bibinfo{pages}{213--216} (\bibinfo{year}{2008}).

\bibitem{szabo2007evolutionary}
\bibinfo{author}{Szab{\'o}, G.} \& \bibinfo{author}{F{\'a}th, G.}
\newblock \bibinfo{title}{Evolutionary games on graphs}.
\newblock \emph{\bibinfo{journal}{Phys. Rep.}} \textbf{\bibinfo{volume}{446}},
  \bibinfo{pages}{97--216} (\bibinfo{year}{2007}).

\bibitem{hauert2004spatial}
\bibinfo{author}{Hauert, C.} \& \bibinfo{author}{Doebeli, M.}
\newblock \bibinfo{title}{Spatial structure often inhibits the evolution of
  cooperation in the snowdrift game}.
\newblock \emph{\bibinfo{journal}{Nature}} \textbf{\bibinfo{volume}{428}},
  \bibinfo{pages}{643--646} (\bibinfo{year}{2004}).

\bibitem{ohtsuki2006simple}
\bibinfo{author}{Ohtsuki, H.}, \bibinfo{author}{Hauert, C.},
  \bibinfo{author}{Lieberman, E.} \& \bibinfo{author}{Nowak, M.~A.}
\newblock \bibinfo{title}{A simple rule for the evolution of cooperation on
  graphs and social networks}.
\newblock \emph{\bibinfo{journal}{Nature}} \textbf{\bibinfo{volume}{441}},
  \bibinfo{pages}{502--505} (\bibinfo{year}{2006}).

\bibitem{roca2009effect}
\bibinfo{author}{Roca, C.~P.}, \bibinfo{author}{Cuesta, J.~A.} \&
  \bibinfo{author}{S{\'a}nchez, A.}
\newblock \bibinfo{title}{Effect of spatial structure on the evolution of
  cooperation}.
\newblock \emph{\bibinfo{journal}{Phys. Rev. E}} \textbf{\bibinfo{volume}{80}},
  \bibinfo{pages}{046106} (\bibinfo{year}{2009}).

\bibitem{gomez2012evolution}
\bibinfo{author}{G{\'o}mez-Garde{\~n}es, J.}, \bibinfo{author}{Reinares, I.},
  \bibinfo{author}{Arenas, A.} \& \bibinfo{author}{Flor{\'\i}a, L.~M.}
\newblock \bibinfo{title}{Evolution of cooperation in multiplex networks}.
\newblock \emph{\bibinfo{journal}{Sci. Rep.}} \textbf{\bibinfo{volume}{2}},
  \bibinfo{pages}{620} (\bibinfo{year}{2012}).

\bibitem{wang2013interdependent}
\bibinfo{author}{Wang, Z.}, \bibinfo{author}{Szolnoki, A.} \&
  \bibinfo{author}{Perc, M.}
\newblock \bibinfo{title}{Interdependent network reciprocity in evolutionary
  games}.
\newblock \emph{\bibinfo{journal}{Sci. Rep.}} \textbf{\bibinfo{volume}{3}},
  \bibinfo{pages}{1183} (\bibinfo{year}{2013}).

\bibitem{gomez2012evolutionary}
\bibinfo{author}{G{\'o}mez-Garde{\~n}es, J.},
  \bibinfo{author}{Gracia-L{\'a}zaro, C.}, \bibinfo{author}{Flor{\'\i}a, L.~M.}
  \& \bibinfo{author}{Moreno, Y.}
\newblock \bibinfo{title}{Evolutionary dynamics on interdependent populations}.
\newblock \emph{\bibinfo{journal}{Phys. Rev. E}} \textbf{\bibinfo{volume}{86}},
  \bibinfo{pages}{056113} (\bibinfo{year}{2012}).

\bibitem{wang2012probabilistic}
\bibinfo{author}{Wang, B.}, \bibinfo{author}{Chen, X.} \&
  \bibinfo{author}{Wang, L.}
\newblock \bibinfo{title}{Probabilistic interconnection between interdependent
  networks promotes cooperation in the public goods game}.
\newblock \emph{\bibinfo{journal}{J. Stat. Mech.}}
  \textbf{\bibinfo{volume}{2012}}, \bibinfo{pages}{P11017}
  (\bibinfo{year}{2012}).

\bibitem{jiang2013spreading}
\bibinfo{author}{Jiang, L.-L.} \& \bibinfo{author}{Perc, M.}
\newblock \bibinfo{title}{Spreading of cooperative behaviour across
  interdependent groups}.
\newblock \emph{\bibinfo{journal}{Sci. Rep.}} \textbf{\bibinfo{volume}{3}},
  \bibinfo{pages}{2483} (\bibinfo{year}{2013}).

\bibitem{wang2013optimal}
\bibinfo{author}{Wang, Z.}, \bibinfo{author}{Szolnoki, A.} \&
  \bibinfo{author}{Perc, M.}
\newblock \bibinfo{title}{Optimal interdependence between networks for the
  evolution of cooperation}.
\newblock \emph{\bibinfo{journal}{Sci. Rep.}} \textbf{\bibinfo{volume}{3}},
  \bibinfo{pages}{2470} (\bibinfo{year}{2013}).

\bibitem{wang2012evolution}
\bibinfo{author}{Wang, Z.}, \bibinfo{author}{Szolnoki, A.} \&
  \bibinfo{author}{Perc, M.}
\newblock \bibinfo{title}{Evolution of public cooperation on interdependent
  networks: The impact of biased utility functions}.
\newblock \emph{\bibinfo{journal}{EPL}} \textbf{\bibinfo{volume}{97}},
  \bibinfo{pages}{48001} (\bibinfo{year}{2012}).

\bibitem{taylor1978evolutionary}
\bibinfo{author}{Taylor, P.~D.} \& \bibinfo{author}{Jonker, L.~B.}
\newblock \bibinfo{title}{Evolutionary stable strategies and game dynamics}.
\newblock \emph{\bibinfo{journal}{Math. Biosci.}}
  \textbf{\bibinfo{volume}{40}}, \bibinfo{pages}{145--156}
  (\bibinfo{year}{1978}).

\bibitem{ohtsuki2007breaking}
\bibinfo{author}{Ohtsuki, H.}, \bibinfo{author}{Nowak, M.~A.} \&
  \bibinfo{author}{Pacheco, J.~M.}
\newblock \bibinfo{title}{Breaking the symmetry between interaction and
  replacement in evolutionary dynamics on graphs}.
\newblock \emph{\bibinfo{journal}{Phys. Rev. Lett.}}
  \textbf{\bibinfo{volume}{98}}, \bibinfo{pages}{108106}
  (\bibinfo{year}{2007}).

\bibitem{ohtsuki2007evolutionary}
\bibinfo{author}{Ohtsuki, H.}, \bibinfo{author}{Pacheco, J.~M.} \&
  \bibinfo{author}{Nowak, M.~A.}
\newblock \bibinfo{title}{Evolutionary graph theory: breaking the symmetry
  between interaction and replacement}.
\newblock \emph{\bibinfo{journal}{J. Theor. Biol.}}
  \textbf{\bibinfo{volume}{246}}, \bibinfo{pages}{681--694}
  (\bibinfo{year}{2007}).

\bibitem{xxx}
\bibinfo{author}{Santos, M.~D.}, \bibinfo{author}{Dorogovtsev, S.~N.} \&
  \bibinfo{author}{Mendes, J. F.~F.}
\newblock \bibinfo{title}{, in preparation} .

\bibitem{barabasi1999emergence}
\bibinfo{author}{Barab{\'a}si, A.-L.} \& \bibinfo{author}{Albert, R.}
\newblock \bibinfo{title}{Emergence of scaling in random networks}.
\newblock \emph{\bibinfo{journal}{Science}} \textbf{\bibinfo{volume}{286}},
  \bibinfo{pages}{509--512} (\bibinfo{year}{1999}).

\bibitem{kollock1998social}
\bibinfo{author}{Kollock, P.}
\newblock \bibinfo{title}{Social dilemmas: The anatomy of cooperation}.
\newblock \emph{\bibinfo{journal}{Annu. Rev. Sociol.}}
  \bibinfo{pages}{183--214} (\bibinfo{year}{1998}).

\bibitem{traulsen2007stochastic}
\bibinfo{author}{Traulsen, A.}, \bibinfo{author}{Nowak, M.~A.} \&
  \bibinfo{author}{Pacheco, J.~M.}
\newblock \bibinfo{title}{Stochastic payoff evaluation increases the
  temperature of selection}.
\newblock \emph{\bibinfo{journal}{J. Theor. Biol.}}
  \textbf{\bibinfo{volume}{244}}, \bibinfo{pages}{349--356}
  (\bibinfo{year}{2007}).

\bibitem{traulsen2007pairwise}
\bibinfo{author}{Traulsen, A.}, \bibinfo{author}{Pacheco, J.~M.} \&
  \bibinfo{author}{Nowak, M.~A.}
\newblock \bibinfo{title}{Pairwise comparison and selection temperature in
  evolutionary game dynamics}.
\newblock \emph{\bibinfo{journal}{J. Theor. Biol.}}
  \textbf{\bibinfo{volume}{246}}, \bibinfo{pages}{522--529}
  (\bibinfo{year}{2007}).

\end{thebibliography}


\begin{acknowledgments}
This work was partially supported by the FCT project PTDC/MAT/114515/2009 and the FET proactive IP
project MULTIPLEX number 317532 and the fellowship grant SFRH/BPD/90936/2012. 
\end{acknowledgments}



\newpage

\setcounter{equation}{0}
\setcounter{figure}{0}
\setcounter{section}{0}

\renewcommand{\theequation}{S\arabic{equation}}
\renewcommand{\thefigure}{S\arabic{figure}}
\renewcommand{\thesection}{S\,\Roman{section}}

\begin{center}
\bf{SUPPLEMENTARY \vspace{14pt} INFORMATION}  
\end{center}

In the main text we focused on the case of $\beta=1$. In particular, for this value of $\beta$, we considered a fully connected network organized in two layers. In this Supplementary Information we provide additional results for other $\beta$ values for this easily treatable network, which demonstrate that the value $\beta=1$ is representative. We present the rate equations in the specific case $p=1$, for which the results in the steady state prove to be independent of $\beta$. Furthermore, we show detailed numerical results for $p<1$, which depend on the particular $\beta$ value chosen. 

\section{{\small Well-mixed populations: results for various $\beta$ values}}

For $p=1$, Eqs. (5) and (6) in the main text simplify to
\begin{equation}
\label{eq::gradPD}
\dot{x}_{\ms{\text{PD}}}= x_{\ms{\text{PD}}} \left( 1 - x_{\ms{\text{PD}}} \right) \tanh \left[ \frac{\beta}{2} \left( f_{\ms{\text{C,PD}}} - f_{\ms{\text{D,PD}}} \right) \right]
\end{equation}
\begin{equation}
\label{eq::gradSG}
\dot{x}_{\ms{\text{SG}}}= x_{\ms{\text{SG}}} \left( 1 - x_{\ms{\text{SG}}} \right) \tanh \left[ \frac{\beta}{2} \left( f_{\ms{\text{C,SG}}} - f_{\ms{\text{D,SG}}} \right) \right]
\end{equation}
The steady state solution of Eqs.~(\ref{eq::gradPD}) and (\ref{eq::gradSG}), assuming $\dot{x}_{\ms{\text{PD}}}=0$ and $\dot{x}_{\ms{\text{SG}}}=0$, is independent of $\beta$. 

For $p<1$, on the other hand, Eqs. (5) and (6) in the main text cannot be simplified in the same way as Eqs.~(\ref{eq::gradPD}) and (\ref{eq::gradSG}), and their steady state solutions depend on the chosen $\beta$ value. In Supplementary Fig.~\ref{newfigSI} we present the fraction of cooperations in the steady state for three $\beta$ values, $\beta=0.1$, $1.0$, and $10$. Here the diagrams for $\beta=1.0$ are represented from Fig.~6 in the main text. 

Comparing the results shown for the various $\beta$ values, we observe that the final level of cooperation shows the same qualitative behavior discussed in the main text regardless of the $\beta$ value. That is, the fraction of cooperators monotonously decreases with decreasing $p$ in the SG layer, and there is an optimal $p$ value for the level of cooperation in the PD layer. As $\beta$ increases from $\beta=0.1$, there is a shift in that optimal $p$ value: for $\beta=0.1$ it is close to $p=0.9$, while on the other hand for $\beta=10$ it is close to $p=0.25$. That is, the optimal $p$ value for the level of cooperation in the PD layer decreases for increasing $\beta$.

The results are also independent of the $\beta$ value adopted for $p=0$: the population always reaches full defection in both layers.

\begin{figure}
\includegraphics[scale=0.6]{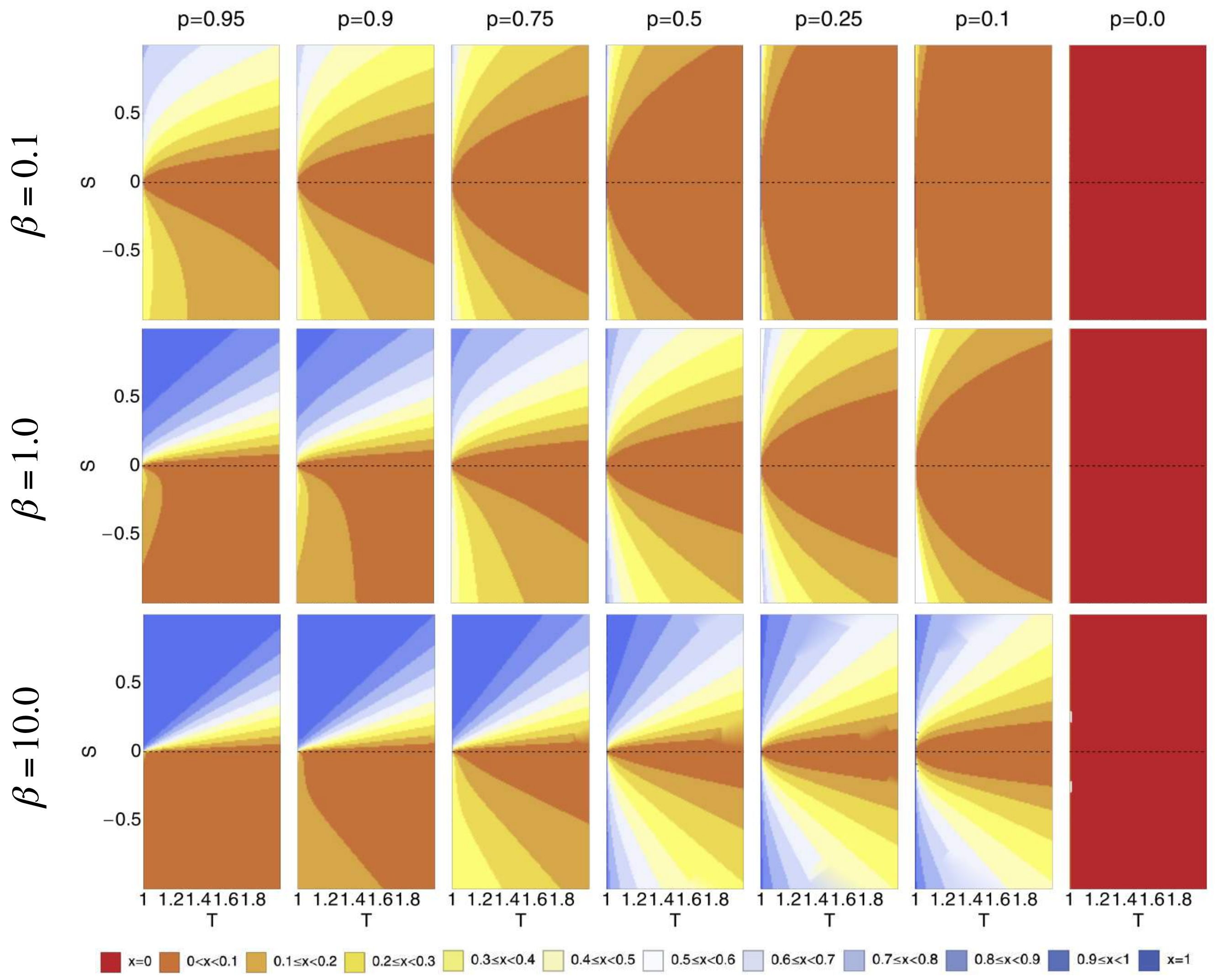}
\caption{\textbf{Final fraction of cooperators on each layer of the well-mixed population for various $\beta$ values.} Contour plots show the final fraction of cooperators $x$ on both the PD $\left( S<0\right)$ and SG $\left( S>0\right)$ layers as a function of $T$ and $S$, for $\beta=0.1$ (first row), $\beta=1$ (second row) and $\beta=10$ (third row). The results for $\beta=1$ are reproduced from the main text for easier comparison. The dashed line separates the results for the SG layer (upper quadrants) from the results for the PD layer (lower quadrants). Red corresponds to full defection, blue corresponds to full cooperation. \label{newfigSI}}
\end{figure}

\end{document}